\documentclass[11pt,numbers=enddot,a4paper]{scrartcl}

\usepackage{amsthm,amssymb,amsmath}
\usepackage{wasysym,latexsym,mathtools,bm}
\usepackage{graphicx,psfrag,color}
\usepackage[ansinew]{inputenc}
\usepackage[USenglish]{babel}

\usepackage{fourier} 
\usepackage[scaled=0.875]{helvet} 
\usepackage{setspace}
\usepackage{microtype} 
\usepackage{lettrine} 
\usepackage{paralist} 

\usepackage{abstract} 

\usepackage{booktabs}
\usepackage[flushmargin,multiple]{footmisc}
\usepackage[comma]{natbib}
\usepackage[top=3.5cm,textheight=23.5cm,textwidth=15cm,footskip=1.5cm]{geometry} 

\usepackage{fancyhdr} 
\usepackage{array}
\newcolumntype{L}[1]{>{\raggedright\let\newline\\\arraybackslash\hspace{0pt}}m{#1}}
\newcolumntype{C}[1]{>{\centering\let\newline\\\arraybackslash\hspace{0pt}}m{#1}}
\newcolumntype{R}[1]{>{\raggedleft\let\newline\\\arraybackslash\hspace{0pt}}m{#1}}

\newfont{\euro}{eurorm scaled \magstephalf}  

\newcommand{\N}{\mathbb{N}}

\newcommand{\R}{\mathbb{R}}

\newcommand{\E}{\textup{E}}

\newcommand{\QED}{\hfill Q.E.D.\medskip}

\newtheorem{theo}{Theorem}
\newtheorem{prop}{Proposition}

\newtheorem{defn}{Definition}

\setkomafont{title}{\rm}%
\setkomafont{section}{\Large\centering\rm}%
\setkomafont{subsection}{\large\centering\rm}%
\setkomafont{subsubsection}{\centering\rm}

\begin{document}
\title{\textbf{A Note on Bayesian Rationality and \\
Correlated Equilibrium}\footnote{\hspace{1ex}I am very grateful to Bob Aumann, Harriet Bergmann, Steve Brams, Rainer Dyckerhoff, Andrés Perea, Mathias Risse and Rainer Schüssler for their helpful comments on the manuscript and our valuable discussions.}}
\author{Gabriel Frahm\thanks{\hspace{1ex}Phone: +49 40 6541-2791, e-mail: frahm@hsu-hh.de.\vspace{.5em}}\\
Helmut Schmidt University\\
Department of Mathematics/Statistics\\
Chair for Applied Stochastics and \\
Risk Management}


\maketitle
\thispagestyle{fancy}

\begin{abstract}
\noindent Bayesian rationality in strategic games presumes that it is possible to translate strategic uncertainty into imperfect information. Correlated equilibrium is guided by the idea that players are Bayes rational, have a common prior, and choose their strategies independently. I show that an essential condition for Bayesian rationality is violated in every game with imperfect information. Moreover, without strategic uncertainty, players cannot choose their strategies independently. This means strategic independence requires strategic uncertainty. If we distinguish between strategic certainty and uncertainty, we are able to explain both the existence of the cooperative and the noncooperative solution of the prisoner's dilemma.
\end{abstract}

\smallskip


\newpage


\begin{quote}
``\textsl{The logical roots of game theory are in Bayesian decision theory. Indeed, game theory can be viewed as an extension of decision theory (to the case of two or more decision-makers), or as its essential logical fulfillment. Thus, to understand the fundamental ideas of game theory, one should begin by studying decision theory.}''
\end{quote}
\hfill\citet[p.~5]{Myerson1991}

\section{Bayesian Rationality in Games}

\lettrine[lines=3,nindent=0pt]{T}{his} note builds on the foundations of subjectivistic decision theory, i.e., of rational choice under uncertainty. Thus, it seems worth recapitulating the basic theory, before going into the details. This section presents typical assumptions about (i) the structure of the decision problem, (ii) the consistency of the decision maker's preferences, and (iii) his behavior or---in the context of game theory---the behavior of the players. A nice overview of subjectivistic expected-utility theories can be found in \citet{Fishburn1981}. Readers familiar with this topic must forgive me and may skip this section.

\citet{AD2009} distinguish between games against nature and strategic games. This suggests that the principles of rational choice hold irrespective of whether we suppose that there is only a single decision maker faced with nature or a number of players competing with each other in a situation of conflict. Hence, each decision maker may be considered a player and vice versa. Further, no distinction is made between cooperative and noncooperative strategic games, since every ``noncooperative'' game can lead to cooperation---and thus turn into a ``cooperative'' game---if this is in the interests of each player \citep{Selten2001}. This means cooperation is viewed as a possible result but not as a prerequisite of a strategic game.

\subsection{Games against Nature}\label{Sec.: Games against Nature}

The following exposition is based on \citet{Fishburn1981} and \citet{Savage1954}. The state space of the decision problem is denoted by $\Omega$. It is assumed that $\Omega$ is nonempty. Each element $\omega\in\Omega$ represents a state of nature or state of the world. Let $\mathcal{F}$ be a nonempty set of subsets of $\Omega$ such that $\Omega\in\mathcal{F}$, $F\in\mathcal{F}\Rightarrow\Omega\setminus F\in\mathcal{F}$, and $F_1,F_2,\ldots\in\mathcal{F}\Rightarrow\bigcup_{i\in\N}F_i\in\mathcal{F}$. Hence, $\mathcal{F}$ is a $\sigma$-algebra and each element of $\mathcal{F}$ is referred to as an event. The event $F\in\mathcal{F}$ is said to happen if and only if some state $\omega\in F$ obtains. The tuple $(\Omega,\mathcal{F})$ represents a measurable space. A function $p\!:\mathcal{F}\rightarrow[0,1]$ is said to be a probability measure if and only if $p(\Omega)=1$ and $p\big(\bigcup_{i\in\N}F_i\big)=\sum_{i\in\N}p(F_i)$ for all mutually disjoint events $F_1,F_2,\ldots\in\mathcal{F}$. It is supposed that the decision maker has a subjective probability measure $p$, i.e., a \textit{prior}, so that $p(F)$ represents his prior probability of $F\in\mathcal{F}$. This leads to a probability space $(\Omega,\mathcal{F},p)$. Moreover, there exists a nonempty set $\mathcal{N}$ of null events, which consists of all $F\in\mathcal{F}$ with $p(F)=0$. The decision maker does not believe that any event $F\in\mathcal{N}$ will happen and thus he may ignore every null event. The null events are called negligible, whereas each event $F\in\mathcal{F}\setminus\mathcal{N}$ is said to be substantial.

\citet[][p.~141]{Fishburn1981} notes that, ``states [\ldots] lead to specific consequences that depend on the course of action adopted by the individual.'' In addition, he writes that, ``[\ldots] the occurrence of one consequence precludes the occurrence of any other consequence.'' Thus, consider a nonempty set $C$ of consequences. Throughout this work, it is assumed that $C$ is a subset of $\R^n$ with $n\in\N$. Let $C^\Omega$ be the set of all $\mathcal{F}$-measurable functions from $\Omega$ to $C$. The decision maker chooses a certain action, which leads to one and only one $s\in C^\Omega$, whereas each state of nature, $\omega$, leads to a specific consequence $s(\omega)\in C$. According to \citet[][Chapter 2.5]{Savage1954}, ``If two different acts had the same consequences in every state of the world, there would [\ldots] be no point in considering them two different acts at all.'' This means in a game against nature it is not necessary to distinguish between the action of the decision maker and the associated element of $C^\Omega$. Hence, we may call each element of $C^\Omega$ a Savage act \citep[][p.~143,160]{Fishburn1981}.

\citet[][p.~15]{Savage1954} treats an important aspect of his subjectivistic approach to rational choice:
\begin{quote}
``\textsl{The argument might be raised that the formal description of decision [\ldots] seems inadequate because a person may not know the consequences of the acts open to him in each state of the world.}''
\end{quote}
Savage argues that it should always be possible to cut each element of $\Omega$ into pieces, i.e., to dissect every potential source of uncertainty, until every state of nature, $\omega$, leads to one and only one consequence $s(\omega)$.\footnote{See the omelet example in \citet[][pp.~13--15]{Savage1954}.} The Savage act, $s$, may depend on the decision, but once the action of the decision maker and the state of nature, $\omega$, are fixed, the decision maker cannot be uncertain about $s(\omega)$. This means he might not know $\omega$, but if somebody tells him $\omega$, he certainly knows $s(\omega)$, i.e., the consequence of his decision that occurs if $\omega\in\Omega$ obtains. This fundamental assumption can be made even more precise as follows:
\begin{enumerate}
  \item[(i)] The state space, $\Omega$, is properly specified. This means the decision maker has a \emph{unique} conjecture about the potential consequences of each possible action.
  \item[(ii)] His conjectures are always correct, i.e., he \emph{knows} the potential consequences.
\end{enumerate}
The above distinction between belief and knowledge seems to be of minor importance in games against nature, but it turns out to be essential when we discuss strategic games.

Similar arguments hold for the set of possible actions. Since the actions are mutually exclusive, the action set of the decision maker should contain every available policy. For example, if we consider a game against nature in its extensive form, where one decision can lead to another, we should take all sequential decision rules into account that can be adopted by the player. Nonetheless, in the normal form of the game, there is no time dimension and thus, at least in a formal sense, actions and consequences cannot be associated with any point in time.

The observation that $s$ must be uniquely determined by the action of the decision maker might seem trivial prima facie, but the devil is in the details. The Savage act $s$ is an element of $C^\Omega$, i.e., the set of all $\mathcal{F}$-measurable functions from $\Omega$ to $C\subseteq\R^n$. Moreover, the decision maker's prior, $p$, forms a quotient space of $C^\Omega$, i.e., a set of equivalence classes. This means two elements of $C^\Omega$ belong to the same equivalence class, i.e., are considered identical by the decision maker, if and only if they are \emph{almost surely} equal. Here, the notion of ``almost surely'' essentially depends on $p$. For example, consider two Savage acts $s,t\in C^\Omega$ and suppose that there are two players with different priors. Then the Savage acts $s$ and $t$ could be identical from the perspective of Player 1. This means $s$ and $t$ differ only with respect to those events that are negligible for Player 1. By contrast, the Savage acts $s$ and $t$ might be different for Player 2 because they differ at least with respect to one event that is substantial from his point of view. Hence, the $L^0$ space of the decision maker essentially depends on his individual prior.

It is assumed that the decision maker has a measurable utility function $u\!:\,\R\rightarrow\R$ such that
\[
\E\big(u(s)\big) := \int_\Omega u\big(s(\omega)\big)\,p(d\omega) < \infty
\]
for all $s\in C^\Omega$. The Savage act $s$ is preferred to $t$, i.e., $s\succ t$, if and only if $\E\big(u(s)\big)>\E\big(u(t)\big)$. Hence, the utility function $u$ induces an asymmetric weak order $\succ$ on $C^\Omega$, i.e., $\succ$ is a strict preference relation such that $s\succ t\Rightarrow\neg(t\succ s)$ and $r\succ t\Rightarrow(r\succ s\,\vee\,s\succ t)$ for all $r,s,t\in C^\Omega$. It can easily be extended to a weak order $\succeq$ \citep[][p.~145]{Fishburn1981}. Now, let $\mathcal{S}$ be a nonempty subset of $C^\Omega$. This may be seen as the set of Savage acts that are available to the decision maker. He is considered rational if and only if he chooses an optimal act $s^*\in\mathcal{S}$. This means there must be no other Savage act $s\in\mathcal{S}$ such that $\E\big(u(s)\big)>\E\big(u(s^*)\big)$. This represents a behavioral assumption of rational choice---besides our given assumptions about structure and consistency.

So far we have assumed that the prior, $p$, and the utility function, $u$, are given. Hence, the preference relation $\succ$ has been obtained as a \emph{result} of $p$ and $u$. Subjectivistic decision theory usually goes the other way around \citep{deFinetti1937,Fishburn1981,Ramsey1931,Savage1954}. This means one starts with some structural assumptions and basic requirements regarding the consistency of some \emph{given} preference relation. Then he shows that the given preferences can be represented by a subjective probability measure $p$ and a utility function $u$ \citep{Fishburn1981}. Hence, a rational decision maker acts \emph{as if} he would maximize his subjective expected utility. Put another way, we only pretend that he knows his prior, utility function, and the available Savage acts, but we do not require that the decision maker calculates his expected utility \emph{de facto}. This means the subjectivistic approach does not try to explain \emph{how} a rational decision maker comes to his conclusions, but it claims that his choice \emph{is} always rational (for whatever reasons) in the specific sense of the theory. This point is mentioned because it seems to be helpful for understanding the particular approach to strategic conflict chosen by \citet{Aumann1987}.

In the Bayesian framework, the decision maker is equipped with a measurable partition $\mathcal{I}$ of $\Omega$, i.e., a set of nonempty and mutually disjoint events whose union equals $\Omega$. It is supposed that $\mathcal{I}$ is finite only for the sake of simplicity but without loss of generality. In the following, $\mathcal{I}$ is referred to as the private information partition of the decision maker. This means he knows which event $I\in\mathcal{I}$ happens and thus considers each event $F\subseteq\Omega\setminus I$ negligible. More precisely, after receiving his private information, he replaces the prior, $p$, by the posterior $p(\cdot\,|\,I)$ with $p(F\,|\,I)\propto p(I\,|\,F)\,p(F)$ for all $F,I\in\mathcal{F}\setminus\mathcal{N}$. Hence, we have that $p(F\,|\,I)=0$ for all $F,I\in\mathcal{F}\setminus\mathcal{N}$ with $F\cap I=\emptyset$. From the decision maker's point of view, each substantial event $I\in\mathcal{I}$ leads to a new probability space where the posterior $p(\cdot|I)$ represents the subjective probability measure. In particular, the posterior forms a new $L^0$ space, i.e., two Savage acts $s,t\in\mathcal{S}$ are considered identical if and only if they coincide except for the events that are negligible \emph{a posteriori}.

Whenever the decision maker knows that some substantial event $I\in\mathcal{I}$ happens, he chooses a \emph{restricted} Savage act $s_I\in C^\Omega$ \citep[p.~160]{Fishburn1981}.\footnote{If $I$ is negligible, the choice of the decision maker that is based on $I$ is negligible, too.} More precisely, $s_I$ is a Savage act that is unique except for the $p(\cdot\,|\,I)$-null events. His particular choices lead to a Savage act $s\in\mathcal{S}$ such that $s(\omega)=s_I(\omega)$ for each substantial event $I\in\mathcal{I}$ and state of nature $\omega\in\Omega$. It is supposed that each Savage act $s\in\mathcal{S}$ can be constructed in this way. The choice of the decision maker is indicated by an action $a\in A$ with $|A|>1$. Throughout this work, it is assumed that the action set $A$ is a subset of $\R$. By acting on the basis of private information, i.e., choosing an action $a\in A$ for each substantial event $I\in\mathcal{I}$, the decision maker creates a $\sigma(\mathcal{I})$-measurable function from $\Omega$ to $A$, where $\sigma(\mathcal{I})$ denotes the $\sigma$-algebra generated by his information partition $\mathcal{I}$.\footnote{This means the given function is constant over each element of $\mathcal{I}$.} This function is the decision maker's strategy against nature, whereas the corresponding Savage act $s$ can be viewed as nature's ``strategy'' against the decision maker. He considers his strategy a control variable and nature's strategy a state variable. This means the decision maker can deliberately choose any strategy that can be constructed on the basis of his private information partition and action set, but he must accept the potential consequences, i.e., the associated ``response'' of nature to his strategy.

The aforementioned arguments require that $s_I$ is \emph{uniquely} determined by the choice of the decision maker on the basis of $I$.\footnote{In fact, the following arguments are similar to those that have already been discussed above for the case $I=\Omega$.} In particular, $s_I$ must not depend on a choice that is made on the basis of another substantial event $J\in\mathcal{I}$. More precisely, since the state space is properly specified, the decision maker has a unique conjecture about the potential consequences of his decision. Let $\psi(I,a)$ be the conjecture of the decision maker---given that he knows that some substantial event $I\in\mathcal{I}$ happens and chooses the action $a\in A$ on the basis of $I$. \citet[][p.~15]{Savage1954} presumes that the decision maker is always \emph{right}. Hence, for each substantial event $I\in\mathcal{I}$ and action $a\in A$, we have that $\psi(I,a)=s$ except for the $p(\cdot\,|\,I)$-null events, where $s$ denotes the \emph{true} response of nature to the strategy of the decision maker. This basic assumption is referred to as the Bayes condition.

Table \ref{Tab.: Angels and Demons} contains a game against nature, i.e., Angels \& Demons, in which this condition is violated. Here the decision maker's payoff---given that some event $I\in\mathcal{I}$ happens---does not depend \emph{only} on his choice on the basis of $I$; it depends also on the decision that he would have made otherwise. This means the response of the omniscient being---to the action, $a$, that the decision maker chooses on the basis of $I$---is not a function of $I$ and $a$ in the proper mathematical sense. I will come back to this argument when analyzing Bayesian rationality in strategic games.

\begin{table}
  \centering
  \caption{Angels \& Demons}\label{Tab.: Angels and Demons}
  \begin{tabular}{l}
  \toprule
  \begin{minipage}{12cm}
  \footnotesize{An omniscient being flips a coin. If the outcome is heads, he plays the role of an angel and offers the decision maker the choice to lead either an honest or a dishonest life. The decision maker is rewarded with a fortune if he decides to be honest and otherwise he gets nothing. If the outcome is tails, the omniscient being reveals himself as a demon. In this case, he pays a fortune if the decision maker decides to be dishonest, but otherwise he is left with nothing. The omniscient being knows whether the decision maker would have decided to be honest or dishonest in either case. Thus, if heads obtains, he pays the amount only if the decision maker would have also decided to be honest in case of tails. Conversely, if tails obtains, the decision maker is rewarded only if he would have also decided to be dishonest in case of heads.}
  \end{minipage}\\
  \bottomrule
  \end{tabular}
\end{table}

The given theoretical approach requires us to split the whole procedure of rational choice notionally into three steps:
\begin{enumerate}
  \item The decision maker knows that some substantial event $I\in\mathcal{I}$ happens.
  \item He chooses an action $a\in A$ on the basis of $I$.
  \item His action leads to a restricted Savage act $s_I\in C^\Omega$.
\end{enumerate}
This decomposition goes beyond the scope of probability theory. Measure theory deals with functions that are considered to be \emph{given}. This means the measure-theoretic viewpoint reflects the situation after the decision maker has made up his mind and so it is \emph{ex post}. By contrast, here we want to analyze the procedure of rational choice, where the decision maker starts with the first step and ends up with the third step. Hence, we must consider the situation before the decision has been made and so the decision-theoretic viewpoint is \emph{ex ante}. By distinguishing between the different points of view we can avoid serious misunderstandings that can arise when discussing situations of strategic conflict, which is done in the following sections.\footnote{The different points of view have nothing to do with the notion of ``prior'' and ``posterior'' in Bayesian decision theory.}

A decision maker is said to be Bayes rational if and only if he chooses an optimal Savage act, given his private information, i.e., one that maximizes his conditional expected utility $\E\big(u(s_I)\,|\,I\big)$ for each substantial event $I\in\mathcal{I}$. More precisely, his action $a\in A$ must lead to a restricted Savage act $s^*_I$ such that $\E\big(u(s^*_I)\,|\,I\big)\geq\E\big(u(s_I)\,|\,I\big)$ for all Savage acts $s_I$ that are available if some substantial event $I\in\mathcal{I}$ happens. Let $s^*\in\mathcal{S}$ be the Savage act of a Bayes-rational decision maker and $s\in\mathcal{S}$ another Savage act. From the law of total expectation it follows that
\[
\E\big(u(s^*)\big) = \E\Big(\E\big(u(s^*_I)\,|\,I\big)\Big) \geq \E\Big(\E\big(u(s_I)\,|\,I\big)\Big) = \E\big(u(s)\big).
\]
We conclude that every Bayes-rational decision maker is rational. Conversely, each rational decision maker is Bayes rational. Otherwise, there would exist a substantial event $I\in\mathcal{I}$ where his action is suboptimal and so the decision maker could increase his unconditional expected utility by substituting the suboptimal action with an optimal one. Hence, Bayesian rationality and rationality (in the unconditional sense) are just two sides of the same coin---provided the Bayes condition is satisfied. If this condition is violated, as it is in Angels \& Demons (see Table \ref{Tab.: Angels and Demons}), we are no longer able to decompose an optimal strategy into a number of single decisions that may be considered optimal \emph{on their own}. In this case, Bayesian rationality is useless. Nonetheless, we can still apply the broader concept of rationality in the unconditional sense.

\subsection{Strategic Games}\label{Sec.: Strategic Games}

The following model of strategic conflict is introduced by \citet{Aumann1987}. In the subsequent analysis, it is referred to as ``Aumann's model.'' Consider a strategic game with $n\in\N\setminus\big\{1\big\}$ players and let $(\Omega,\mathcal{F})$ be the measurable space of the game.\footnote{In contrast to \citet{Aumann1987}, I do not presume that $\Omega$ is finite.} Each player is equipped with a prior $p_i$, a utility function $u_i$, an action set $A_i\subseteq\R$ with $|A_i|>1$, and a private information partition $\mathcal{I}_i$ ($i=1,2,\ldots,n$).\footnote{In the following, the enumeration ``$i=1,2,\ldots,n$'' will be omitted whenever it seems clear from the context that the corresponding statement applies to each player.} It is assumed that the join $\bigvee_{i=1}^n\mathcal{I}_i$, i.e., the coarsest common refinement of $\mathcal{I}_1,\mathcal{I}_2,\ldots,\mathcal{I}_n$, is measurable. Each player may think of as a game against nature where ``nature'' is represented by the other players. Hence, from the perspective of Player $i$, his own strategy is a control variable, whereas the strategy of every player $j\neq i$ is a state variable. It is always supposed that the strategy of Player $i$, $s_i$, is a $\sigma(\mathcal{I}_i)$-measurable function from $\Omega$ to $A_i$. The corresponding strategies of the other players are given by the $(n-1)$-tuple $s^i$.

In a game against nature, a Savage act is the argument of the utility function of the decision maker. The same holds true in a strategic game, but there the utility of Player $i$ depends on his own strategy \emph{and} the strategies of the other players. This means the Savage act of Player $i$ is not his own strategy, but it consists of the entire strategy tuple $s=\big(s_i,s^i\big)\equiv(s_1,s_2,\ldots,s_n)$. Hence, $u_i(s(\omega))$ represents the utility of Player $i$, given that he applies the strategy $s_i$, the others apply the strategies given by $s^i$, and state $\omega\in\Omega$ obtains.

According to \citet[][p.~6]{Aumann1987}, ``The term `state of the world' implies a definite specification of all parameters that may be the object of uncertainty [\ldots].'' Moreover, ``Conditional on a given $\omega$, everybody knows everything.'' Hence, the players might not know the state of the world, $\omega$, but if $\omega$ were revealed to them, they would certainly know each other's action. This reflects the decision-theoretic framework discussed in Section \ref{Sec.: Games against Nature}. However, some statements in \citet{Aumann1987} could lead to misunderstandings. He points out that ``Nash equilibrium does make sense if one starts by assuming that, for some specified reason, each player knows which strategies the other players are using. But this assumption appears rather restrictive.'' Further, he mentions that ``in our treatment, the players do \emph{not} in general know how others are playing.'' Despite these statements, the structural assumptions that are made by Aumann suggest that every player \emph{knows} the strategies of the others.\footnote{This does not mean that anyone knows the \emph{actions} of the others and so the players may have imperfect information.} In Aumann's model of strategic conflict, Player $i$ deals with $s_i$ and $s^i$, but the latter contains nothing other than the actual strategies of his opponents. Hence, since every player considers the same strategy tuple $s=(s_1,s_2,\ldots,s_n)$ a Savage act, at least it is supposed that the players act \emph{as if} they know each other's strategy.

Moreover, \citet[p.~2]{Aumann1987} states that ``it is common knowledge that all the players are Bayesian utility maximizers, that they are rational in the sense that each one conforms to the Savage theory.'' In the Bayes-Savage framework we do not scrutinize the strategic reasoning of the players. Hence, it is not necessary to assume that the rationality of the players is common knowledge. The same holds for their priors, utility functions, information partitions, etc. In fact, \citet[][p.~10]{Aumann1987} points out that these assumptions only ``aid us in understanding the model, they do not affect the conclusions.'' Throughout this work, we do not suppose that the players have common knowledge---neither of their rationality nor of any other personal characteristics.

\citet[p.~2]{Aumann1987} concludes that,
\begin{quote}
``\textsl{According to the Bayesian view, subjective probabilities should be assignable to every prospect, including that of players choosing certain strategies in certain games. Rather than playing an equilibrium, the players should simply choose strategies that maximize their utilities given their subjective distributions over the other players' strategy choices.}''
\end{quote}

This means strategic uncertainty can be treated like uncertainty in a game against nature.\footnote{This view is also held by \citet{AD2009}.} Put another way, it should always be possible to translate strategic uncertainty into imperfect information. A well-known application of this principle is Harsanyi's (\citeyear{Harsanyi1967}) pioneering work on games with incomplete information. Hence, in Aumann's model of strategic conflict, the players cannot suffer from strategic uncertainty, but in general they have imperfect information. Moreover, he assumes that the players are Bayes rational. This means whenever Player $i$ receives some substantial information, i.e., if some event $I_i\in\mathcal{I}_i$ with $p_i(I_i)>0$ happens, he chooses an optimal action $a^*_i\in A_i$. More precisely, $a^*_i$ is such that $\E_i\big(u_i\big(a^*_i,s^i\big)\,|\,I_i\big)\geq\E_i\big(u_i\big(a_i,s^i\big)\,|\,I_i\big)$ for all $a_i\in A_i$ \citep[][p.~7]{Aumann1987}.\footnote{Following Section \ref{Sec.: Games against Nature}, we could also write ``$a_{i,I_i}$'' and ``$s^i_{I_i}$,'' but this is omitted for notational convenience.} Here, $\E_i(\cdot\,|\,I_i)$ denotes the conditional expectation of Player $i$---given the substantial event $I_i\in\mathcal{I}_i$ and individual prior $p_i$.

The tuple $\big(a_i,s^i\big)$ is a restricted Savage act that Player $i$ can choose on the basis of private information. Here, $a_i$ represents not only an action, it is also considered a \emph{consequence} of the decision of Player $i$ that is made on the basis of $I_i\in\mathcal{I}_i$. As in a game against nature, $s^i$ might depend on $a_i$. More precisely, for each substantial event $I_i\in\mathcal{I}_i$ and action $a_i\in A_i$, Player $i$ has a conjecture $\psi_i(I_i,a_i)$, about the strategies of the other players, which is unique except for the $p_i(\cdot\,|\,I_i)$-null events. Now, the Bayes condition can be expressed as follows:
\begin{enumerate}
  \item[\textbf{BAY}.] Player $i$ is never wrong. This means we have that $\psi_i(I_i,a_i)=s^i$ for each substantial event $I_i\in\mathcal{I}_i$ and action $a_i\in A_i$---except for the $p_i(\cdot\,|\,I_i)$-null events.
\end{enumerate}
According to the notation chosen above, \citet{Aumann1987} makes another implicit assumption:
\begin{enumerate}
  \item[\textbf{INV}.] The conjectures do not depend on the actions of Player $i$. More precisely, we have that $\psi_i(I_i,a'_i)=\psi_i(I_i,a''_i)$ for each substantial event $I_i\in\mathcal{I}_i$ and all actions $a'_i,a''_i\in A_i$.
\end{enumerate}
This invariance property is crucial for the derivation of correlated equilibrium.

Assumption \textbf{BAY} $\wedge$ \textbf{INV} implies that no player has an influence on each other's strategy.\footnote{Player $i$ has no influence on the event $I_i\in\mathcal{I}_i$ that is going to happen. He can only control his \emph{action} based upon $I_i$.} More precisely, if Player $i$ changes his strategy, the other players maintain their strategies. This behavioral assumption is referred to as strategic independence. It is typically justified by the common idea that the strategy choices are independent of each other.\footnote{I am very grateful to Andrés Perea for clarifying this point in a personal communication.} However, this is not to say that the chosen strategies are, in any sense, stochastically independent. Hence, strategic independence must be clearly distinguished from stochastic independence. Table \ref{Tab.: Rendezvous} contains a game in which the strategic-independence assumption is violated.

Aumann's assumption shall be denoted by \textbf{AUM} $:=$ \textbf{BAY} $\wedge$ \textbf{INV}. This looks as if we are going to adopt the measure-theoretic, i.e., \emph{ex-post}, point of view discussed in Section \ref{Sec.: Games against Nature}. The problem is that we want to analyze the strategic behavior of the players and thus we must adopt the decision-theoretic, i.e., \emph{ex-ante}, point of view. If we adopt the \emph{ex-ante} point of view for Player $i$, we cannot at the same time adopt the \emph{ex-post} point of view for any player $j\neq i$. Put another way, if we consider the strategies of the other players \emph{after} they have made their decisions, Player $i$ has already made up his mind, too. This would not serve the purpose. Hence, Assumption \textbf{AUM} implies that the strategies of the other players do not depend \emph{ex ante} on the choice of Player $i$. In the following, I always adopt the \emph{ex-ante} point of view when I compare the available actions or strategies of the players with each other. By contrast, if I consider the given solution of a game, I adopt the \emph{ex-post} point of view.

\begin{table}
  \centering
  \caption{Rendezvous}\label{Tab.: Rendezvous}
  \begin{tabular}{l}
  \toprule
  \begin{minipage}{12cm}
  \footnotesize{At the breakfast table, Mary and Joe decide to go to a restaurant after work. They consider Luigi's Trattoria and Harry's Sports Bar. Joe is in a hurry and asks Mary to reserve a table. After Joe has left the room, Mary notices that they did not agree on a choice and now they have absolutely no possibility of communicating with each other. Mary believes that Joe is a smart guy. This means he will call a restaurant to ask whether she has made a reservation. Hence, independent of her particular choice, he will always come to the right place.}
  \end{minipage}\\
  \bottomrule
  \end{tabular}
\end{table}

For the time being, we may suppose that Assumption \textbf{AUM} is satisfied. Hence, the solution of the game, $s^*=(s^*_1,s^*_2,\ldots,s^*_n)$, must be such that
\[
\E_i\Big(u_i\big(s^*_i,s^{*i}\big)\Big) \geq \E_i\Big(u_i\big(s_i,s^{*i}\big)\Big),\qquad\forall s_i,~i=1,2,\ldots,n\,.\footnote{\citet[][p.~4]{Aumann1987} compares ``$f:=\big(f^1,\ldots,f^n\big)$'' with ``$\big(f^{-i},g^i\big):=\big(f^1,\ldots,f^{i-1},g^i,f^{i+1},\ldots,f^n\big)$.'' Here $f^i$ denotes the strategy of Player $i$ and $g^i$ is a function of $f^i$, i.e., the strategy $g^i$ must be $\sigma(\mathcal{I}_i)$-measurable, too.}
\]
This solution is said to be a \emph{subjective} correlated equilibrium \citep[][p.~14]{Aumann1987}. Moreover, \citet[][p.~7]{Aumann1987} presumes that the players have a common prior, i.e., we have that $p_i(F)=p_j(F)$ for all $F\in\mathcal{F}$ and $i,j=1,2,\ldots,n$. This fundamental assumption goes back to \citet{Harsanyi1967}. It implies that the posteriors of the players can differ only through their private information---that is, their personal \emph{evidence}. Otherwise, the deviations could also be due to their individual \emph{beliefs}, which are expressed by $p_1,p_2,\ldots,p_n$. We conclude that the solution of the game, $s^*$, must be such that
\[
\E\Big(u_i\big(s^*_i,s^{*i}\big)\Big) \geq \E\Big(u_i\big(s_i,s^{*i}\big)\Big),\qquad\forall s_i,~i=1,2,\ldots,n\,,
\]
where $\E(\cdot)$ denotes the expectation of each player based on the common prior. This solution is said to be a correlated equilibrium \citep[p.~4]{Aumann1987}.

Common priors are indispensable if we want to guarantee that probabilistic statements are independent of individual beliefs. Otherwise, statements like ``Strategy $s_1$ is stochastically independent of Strategy $s_2$'' and ``Player 1 chooses Action $a_1$ with probability $\frac12$'' can no longer be made without specifying the underlying probability measure, i.e., the prior, of the corresponding player. This means that if the priors are distinct, it is impossible to characterize the solution of the game by a single profile distribution, i.e., a joint probability distribution of strategies. \citet[][p.~12]{Aumann1987} points out that ``Common priors are explicit or implicit in the vast majority of the differential information literature in economics and game theory.'' For a broad overview of the common-prior assumption in economics see \citet{Morris1995}.

Despite its wide acceptance in economics, the common-prior assumption is the subject of controversial discussion \citep{Aumann1998,Gul1998,Morris1995}. Although Aumann generally supports the common-prior assumption, in \citet[][p.~12]{Aumann1987} he mentions that it ``is not a tautological consequence of the Bayesian approach.'' In fact, \citet[][p.~3]{Savage1954} points out that his personalistic interpretation of probability does ``not deny the possibility that two reasonable individuals faced with the same evidence may have different degrees of confidence in the truth of the same proposition.'' \citet{Morris1995} goes even further and argues that it makes little sense, on the one hand, to allow for individual beliefs and, on the other hand, to impose the common-prior assumption, which postulates that the beliefs of the players are equal. The following results do not require the common-prior assumption. It is supposed only that the players agree about the negligible events, i.e., that their priors are \emph{equivalent}.

\section{Strategic Certainty}

\subsection{A Simple Illustration}

\begin{figure}
  \centering
  \includegraphics[scale=.4]{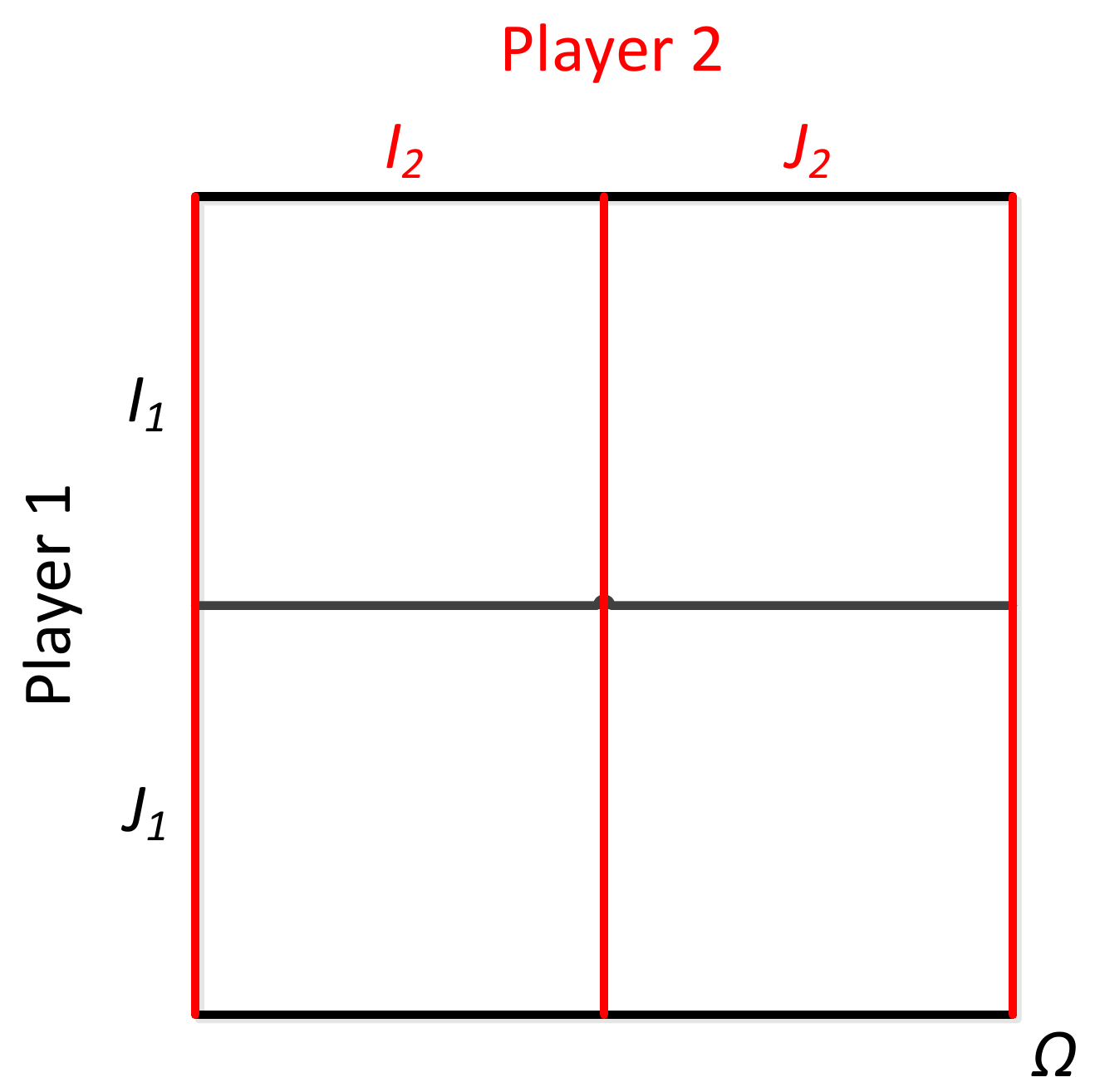}\\[-1ex]
  \caption{State space and information partitions.}\label{Fig.: State space}
\end{figure}

A strategic game is a subtle object. To avoid unnecessary complications, I suppose that there are only two players. The state space of the game, $\Omega$, is depicted in Figure \ref{Fig.: State space}. The event set, $\mathcal{F}$, is the $\sigma$-algebra generated by the partition $\big\{I_1\cap I_2,I_1\cap J_2,J_1\cap I_2,J_1\cap J_2\big\}$ of $\Omega$, and the private information partitions of the players are given by $\mathcal{I}_1=\big\{I_1,J_1\big\}$ and $\mathcal{I}_2=\big\{I_2,J_2\big\}$, respectively. The players assign each element of $\mathcal{F}$---except for $\emptyset$---a positive probability. Further, the action sets correspond to $A_1=A_2=\R$.

Suppose that some state $\omega\in I_1\cap I_2\subset\Omega$ obtains. In this case, Player 1 knows that $I_1$ happens, whereas Player 2 knows that $I_2$ happens. Assume that Player 1 decides to choose Action ``3.'' His action implies a certain strategy of Player 2. The point is that, due to Assumption \textbf{BAY}, the action of Player 1 \emph{uniquely} determines the strategy of Player 2---except for the events that Player 1 considers negligible a posteriori. More precisely, it is assumed that Player 1 \emph{knows} the response of Player 2. For example, if Player 1 chooses ``3,'' Player 2 might choose ``2'' in case of $\omega\in I_2$ and ``1'' if $\omega\in J_2$ obtains. Now, since $\omega\in I_2$ obtains, Player 2 decides to choose ``2.'' In the same way, the choice of Player 2 must lead to a definite strategy of Player 1, which is \emph{known} by Player 2. Player 1 chooses ``3'' in case of $\omega\in I_1$ and, additionally, we can assume that he chooses ``4'' if $\omega\in J_1$ obtains.

The players are Bayes rational and thus Action ``3'' must be an optimal choice for Player 1---provided he knows that the event $I_1$ happens. This means there is no other action $a_1\in\R$ whose (conditional) expected utility is greater than the expected utility of Action ``3.'' In general, Player 1 would have to take into account the response of Player 2 that is associated with each available action $a_1\in A_1$. However, Assumption \textbf{INV} guarantees that Player 2 still chooses Action ``2'' in case of $\omega\in I_2$ and ``1'' if $\omega\in J_2$ obtains, irrespective of whether Player 1 chooses Action ``3'' or something other than ``3'' in case of $\omega\in I_1$. In short, Player 2 will not change his strategy if Player 1 changes his action.

The problem is that Player 1 \emph{cannot choose any other action} if Player 2 sticks with his choice. We have already specified that Player 2 knows that Player 1 chooses Action ``3'' if $\omega\in I_1$ obtains, given that Player 2 chooses Action ``2'' in case of $\omega\in I_2$. Now, if the state $\omega\in I_1\cap I_2$ obtains, it cannot happen that Player 1 chooses ``3'' \emph{and} any other action, simultaneously. Hence, Player 1 is not able to move from ``3'' to anything else---if Player 2 adheres to Action ``2.'' Put another way, \emph{if} Player 1 moves from ``3'' to something else, then Player 2 must change his action as well and thus Assumption \textbf{INV} is violated. The same arguments apply mutatis mutandis to Player 2.

The above arguments work only because we assume that each player \emph{knows} the potential reactions of the other player. In fact, the problem disappears instantly if we allow the conjectures of the players to be \emph{incorrect}. Then Player 2 is ignorant of whatever Player 1 \emph{actually} does and so Player 1 indeed can change his action unilaterally. However, as long as we assume that the players know each other's response, they cannot choose their actions independently.

\begin{figure}
  \centering
  \includegraphics[scale=.32]{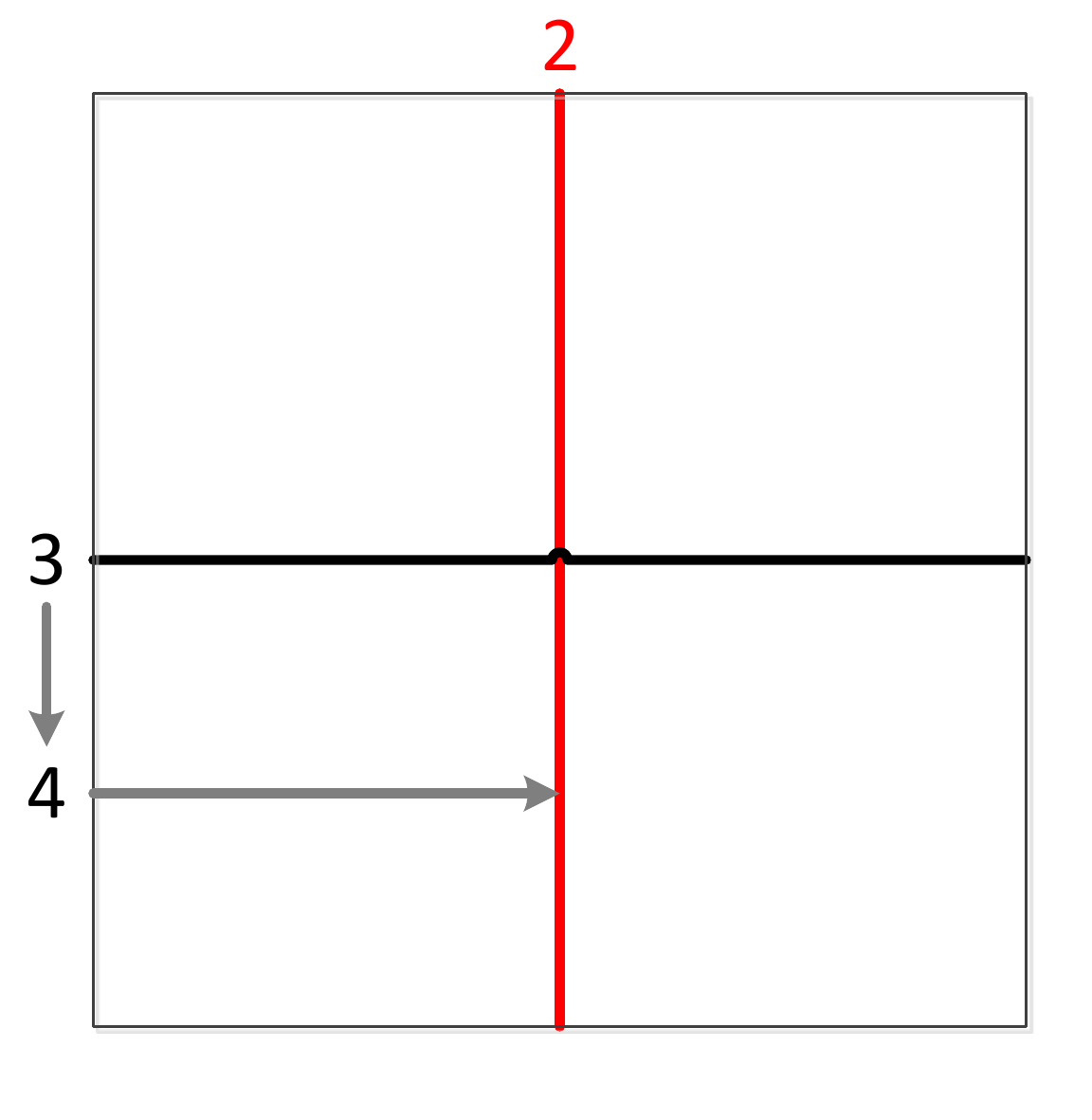}\includegraphics[scale=.32]{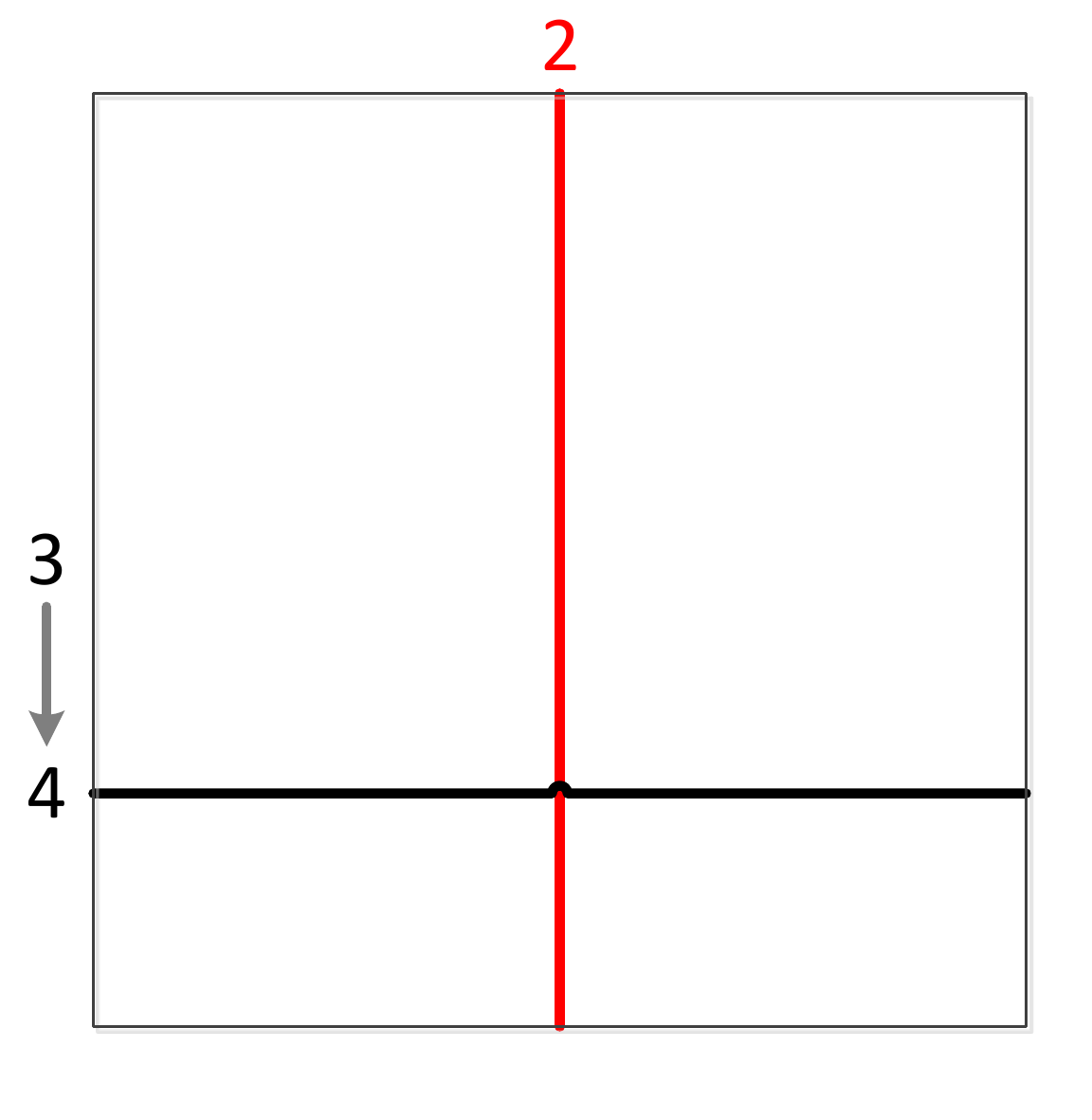}\includegraphics[scale=.32]{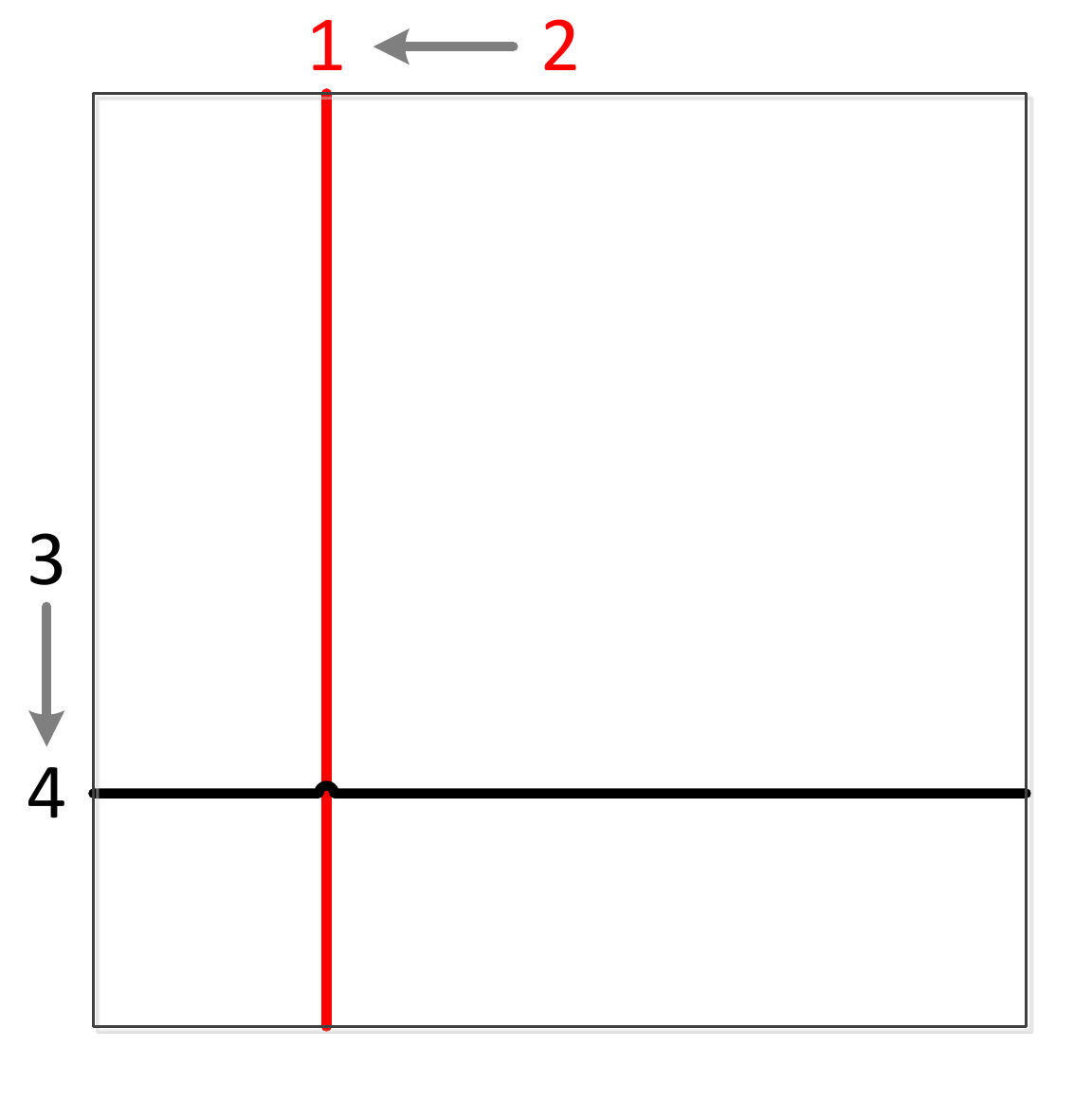}\includegraphics[scale=.32]{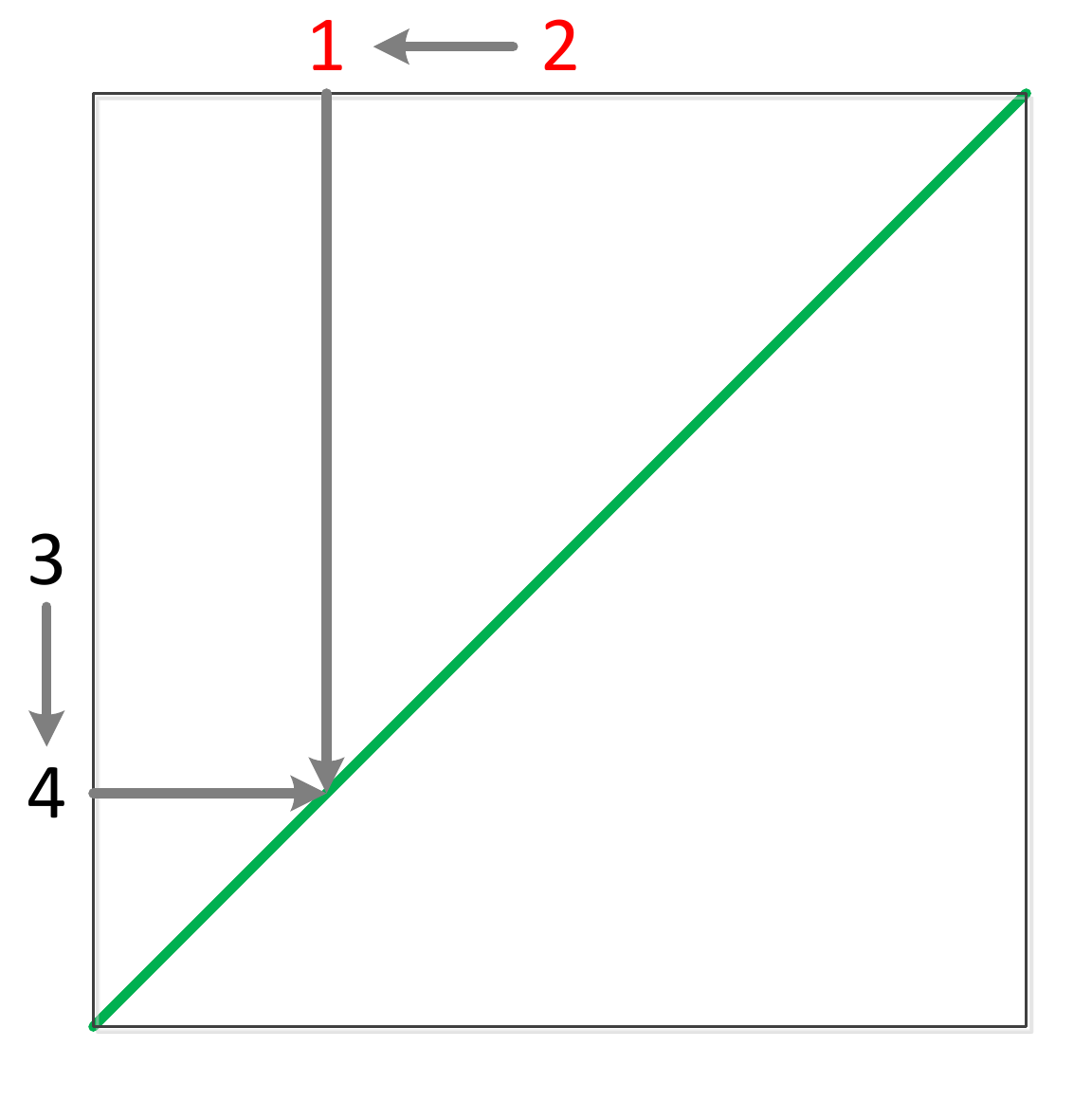}\\[-1ex]
  \caption{Response diagrams for $I_1\cap I_2$.}\label{Fig.: Response diagrams}
\end{figure}

These conclusions could be elusive for some readers, so I would like to illustrate the logic behind my arguments in Figure \ref{Fig.: Response diagrams}. Assumption \textbf{BAY} guarantees that each player has a response function. We may focus, without loss of generality, on the actions that are chosen on the basis of $I_1\in\mathcal{I}_1$ and $I_2\in\mathcal{I}_2$, respectively. The red line in the first diagram indicates the response function of Player 1, i.e., the response of \emph{Player 2} to each possible action of Player 1. It implies that Player 2 will not change his action, ``2,'' if Player 1 moves from ``3'' to ``4.'' By contrast, the response function of Player 2 is given by the black line, which says that Player 1 will not change his action, ``3,'' if Player 2 moves from ``2'' to anything else. Hence, the given response functions reflect Assumption \textbf{INV}. In this case, Player 1 cannot move to ``4.'' This can be seen as follows: If Player 1 moves to ``4,'' he knows that Player 2 chooses ``2,'' but then---according to the response function of Player 2---Player 2 knows that Player 1 chooses ``3.'' This is a contradiction. The same arguments apply to Player 2, i.e., he cannot choose anything other than ``2.'' In fact, the only possible combination of actions is $(3,2)$, i.e., the point of intersection. The problem with this conclusion is twofold: (i) We have assumed that each player can choose an \emph{arbitrary} element of $\R$ and (ii) it makes no sense to restrict the players to a single action, since in this case the question of rational choice becomes trivial.

We could try to resolve this problem by assuming that the black line, i.e., the response function of Player 2, depends on the action of Player 1. More precisely, as is shown in the second diagram, if Player 1 moves from ``3'' to ``4,'' the black line also moves from ``3'' to ``4.'' We can transfer this principle to Player 2, i.e., if Player 2 moves from ``2'' to ``1,'' the same happens with the red line---see the third diagram. If we allow for this possibility, each arbitrary combination of actions within the product space $A_1\times A_2=\R^2$ is possible. However, in this case, the action of Player 2 is not a function, in the proper mathematical sense, of the action of Player 1 and vice versa.\footnote{More precisely, the action of Player 2 is not uniquely determined by the action of Player 1 and vice versa.} This means Assumption \textbf{BAY} is violated and so the given ``response functions'' become meaningless.

As mentioned above, a player can choose only an action that belongs to an intersection point. Moreover, since we assume that both players are free to choose everything from $\R$, the response functions must be congruent and bijective in $\R^2$. In this case, each player can do whatever he wants to do, but the choice of one player depends on the choice of the other. Put another way, their actions are interconnected; i.e., if one player changes his action the other must change his action, too. This is shown in the fourth diagram, where the green line represents the response functions of \emph{both} players. However, this is only one possible solution. We could also have chosen any other bijection in $\R^2$. In any case, Assumption \textbf{INV} cannot be maintained if Assumption \textbf{BAY} holds true. The following section elaborates the aforementioned insights in a more formal way.

\subsection{The Coherence Principle}

Throughout this section, it is implicitly assumed that $n\in\N\setminus\big\{1\big\}$. Thus, we deal with strategic games. Further, it is supposed only that the priors $p_1,p_2,\ldots,p_n$ are equivalent on the measurable space $(\Omega,\mathcal{F})$, i.e., we have that $p_i(F)=0$ if and only if $p_j(F)=0$ for all $F\in\mathcal{F}$ and $i,j=1,2,\ldots,n$. This means the players agree about the negligible events. This assumption is satisfied if the players have a common prior and, indeed, if there exists an objective prior. Player $i$ is said to have imperfect information if and only if there exists a substantial event $I_i\in\mathcal{I}_i$ such that $0<p_i(I_j\,|\,I_i)<1$ for some $I_j\in\mathcal{I}_j$ with $j\neq i$. Otherwise, he has perfect information.

The following results do not require any behavioral assumption. In particular, it is not assumed that the players are (Bayes) rational unless otherwise stated.

\begin{theo}\label{Th.: BAY -> not INV}
Assumption\/ \textup{\textbf{BAY}} precludes Assumption\/ \textup{\textbf{INV}}.
\end{theo}

\noindent Proof: Since the priors are equivalent, the players share the same set $\mathcal{N}$ of null events. Consider Player $i$ and let $I_i\in\mathcal{I}_i$ be some substantial event. Then, for every other player $j\neq i$, we can find an event $I_j\in\mathcal{I}_j$ such that $I_i\cap I_j\in\mathcal{F}\setminus\mathcal{N}$, which implies that $I_j\in\mathcal{F}\setminus\mathcal{N}$. Player $i$ can choose an arbitrary action $a'_i\in A_i$ on the basis of $I_i$. Suppose that Assumption \textbf{BAY} is satisfied. Hence, the strategy of Player $j$ is uniquely determined by $a'_i$ except for the $p_i(\cdot\,|\,I_i)$-null events. More precisely, Player $j$ chooses an action $a_j\in A_j$ on the basis of $I_j$. Since Player $i$ considers $I_i\cap I_j$ substantial, he \emph{knows} that Player $j$ chooses $a_j$ if he chooses $a'_i$. Due to the same arguments, Player $j$ knows that Player $i$ chooses $a_i\in A_i$ if he chooses $a_j$ on the basis of $I_j$. If we suppose that $a_i\neq a'_i$, we obtain the contradiction $a'_i\Rightarrow a_j\Rightarrow a_i$, i.e., Player $i$ cannot choose $a'_i$. We have assumed that he is free to choose every element from $A_i$ and thus we must have that $a_i= a'_i$. Now, suppose that Player $i$ knows that Player $j$ still chooses $a_j$ if he chooses some action $a''_i\in A_i$ with $a''_i\neq a'_i$. Then we obtain the contradiction $a''_i\Rightarrow a_j\Rightarrow a_i=a'_i$, i.e., Player $i$ cannot choose $a''_i$. Hence, since Player $i$ is free to choose anything from $A_i$, Player $j$ cannot adhere to $a_j$ if Player $i$ chooses $a''_i$ instead of $a'_i$. We conclude that Assumption \textbf{INV} is violated.\QED

The proof of Theorem \ref{Th.: BAY -> not INV} reveals that the players cannot choose their strategies independently if Assumption \textbf{BAY} is satisfied. The next theorem states that the Bayes condition is violated \emph{per se} if any player has imperfect information.

\begin{theo}\label{Th.: Reductio ad absurdum}
In every strategic game with imperfect information, Assumption\/ \textup{\textbf{BAY}} is violated.
\end{theo}

\noindent Proof: Since the priors are equivalent, the players share the same set $\mathcal{N}$ of null events. Assume that Player $i$ has imperfect information. This means there exists a substantial event $I_i\in\mathcal{I}_i$ such that $I_i\cap I_j,I_i\cap J_j\in\mathcal{F}\setminus\mathcal{N}$ with $I_j,J_j\in\mathcal{I}_j$ and $I_j\neq J_j$ for some $j\neq i$. Suppose that Assumption \textbf{BAY} is satisfied. Then every action that Player $i$ chooses on the basis of $I_i$ uniquely determines the actions chosen by Player $j$ both on $I_j$ and $J_j$. Thus, if Player $j$ adheres to his action on $I_j$ but changes his action on $J_j$, Player $i$ must change his action on $I_i$, too. From $I_i\cap I_j,I_i\cap J_j\in\mathcal{F}\setminus\mathcal{N}$ we conclude that $I_j,J_j\in\mathcal{F}\setminus\mathcal{N}$. Assumption \textbf{BAY} implies that the action of Player $i$ on $I_i$ must be uniquely determined by the action of Player $j$ on $I_j$. However, we have already concluded that Player $i$ changes his action on $I_i$ if Player $j$ sticks with his action on $I_j$ but changes his action on $J_j$. Hence, the action of Player $i$ on $I_i$ is not uniquely determined by the action of Player $j$ on $I_j$ and thus Assumption \textbf{BAY} is violated.\QED

Theorem \ref{Th.: Reductio ad absurdum} renders Bayesian rationality in games with imperfect information useless. The problem is that Player $i$ cannot predict the actions of Player $j$ only by taking his action---based on a substantial event $I_i\in\mathcal{I}_i$---into consideration. His action represents a (very small) part of his overall strategy, but the potential actions of Player $j$, i.e., his strategy, depends on the overall strategy of Player $i$---not on the single action that he chooses on the basis of $I_i$. This situation reminds us a bit of Angels \& Demons (see Table \ref{Tab.: Angels and Demons}). More precisely, Player $i$ represents the decision maker and Player $j$ can be viewed as the omniscient being. By contrast, if the players have perfect information, i.e., know each other's reaction, we can apply Theorem \ref{Th.: BAY -> not INV}. In this case, the players cannot choose their strategies independently and thus every Bayes-rational player has to take the responses of the others into account when maximizing his (conditional) expected utility. This is demonstrated below by means of an example.

Theorem \ref{Th.: Reductio ad absurdum} only implies that \emph{Bayesian} rationality cannot be used as a solution concept for strategic games with imperfect information. Nonetheless, we must not conclude that our whole model of strategic conflict becomes useless. Aumann's subjectivistic approach requires only that Player $i$ knows the responses of the others to his own \emph{strategy}. However, the strategies of the others may depend on the strategy of Player $i$. More precisely, Player $i$ has a \emph{unique} conjecture, $\Psi_i(s_i)$, about the responses of the other players to each strategy $s_i$ and we make the following additional assumption:
\begin{enumerate}
  \item[\textbf{SAV}.] We have that $\Psi_i(s_i)=s^i$ for each strategy $s_i$.
\end{enumerate}

The fact that the conjecture of Player $i$ is uniquely determined by his own strategy precisely reflects the aforementioned idea that the state space, $\Omega$, is properly specified. Assumption \textbf{SAV} can be considered an unconditional version of Assumption \textbf{BAY}. Hence, according to \citet[][p.~15]{Savage1954}, we assume that the decision maker, i.e., Player $i$, is never wrong. This assumption is referred to as strategic certainty. Hence, each strategy $s_i$ implies some response $s^i$. This may be expressed in a more compact way as ``$s_i\Rightarrow s^i$.''

In the remainder of this section, we shall analyze the question of whether we can maintain Aumann's model of strategic conflict without resorting to Bayesian rationality. The analogue to Assumption \textbf{INV} reads as follows:
\begin{enumerate}
  \item[\textbf{FIX}.] We have that $\Psi_i(s'_i)=\Psi_i(s''_i)$ for all strategies $s'_i$ and $s''_i$.
\end{enumerate}
Assumption \textbf{SAV} $\wedge$ \textbf{FIX} implies that the other players maintain their strategies if Player $i$ changes his strategy. This is nothing other than the strategic-independence assumption discussed in Section \ref{Sec.: Strategic Games}. Under these circumstances the players could be considered rational---but not \emph{Bayes} rational---if and only if their strategies belong to a subjective correlated equilibrium. Moreover, if the players have a common prior, the solution of the game must be a correlated equilibrium.

The next theorem denies the coexistence of Assumption \textbf{SAV} and Assumption \textbf{FIX}. This means Assumption \textbf{SAV} $\wedge$ \textbf{FIX} cannot be satisfied.

\begin{theo}\label{Th.: Main Theorem}
Assumption\/ \textup{\textbf{SAV}} precludes Assumption\/ \textup{\textbf{FIX}}.
\end{theo}

\noindent Proof: If Assumption \textbf{SAV} is satisfied, each strategy of Player $i$ leads to one and only one strategy of Player $j$. Suppose that Player $i$ applies some strategy $s'_i$ and Player $j$ responds with the strategy $s_j$. Analogously, the strategy of Player $j$, $s_j$, leads to one and only one response, $s_i$, of Player $i$. If we suppose that $s_i\neq s'_i$ we obtain the contradiction $s'_i\Rightarrow s_j\Rightarrow s_i$, i.e., Player $i$ cannot apply $s'_i$. Indeed, he \emph{can} choose every $\sigma(\mathcal{I}_i)$-measurable function from $\Omega$ to $A_i$ and so we must have that $s_i=s'_i$. Let Assumption \textbf{FIX} be satisfied, too. This means Player $j$ maintains his strategy, $s_j$, if Player $i$ moves from $s'_i$ to $s''_i\neq s'_i$. Now, we obtain the contradiction $s''_i\Rightarrow s_j\Rightarrow s_i=s'_i$, i.e., Player $i$ is not able to apply $s''_i$. In fact, he \emph{is} allowed to apply every strategy that can be constructed on the basis of his private information partition and action set. We conclude that Assumption \textbf{FIX} is violated.\QED

Hence, if Assumption \textbf{SAV} is satisfied, there exist two different strategies $s'_i$ and $s''_i$ such that $\Psi_i(s'_i)\neq \Psi_i(s''_i)$, where the conjectures $\Psi_i(s'_i)$ and $\Psi_i(s''_i)$ reflect the \emph{true} responses of the other players to the strategies $s'_i$ and $s''_i$ of Player $i$. Hence, strategic certainty crowds out strategic independence or, equivalently, strategic independence requires strategic uncertainty.

Now, we maintain Assumption \textbf{SAV} but drop Assumption \textbf{FIX}. This means we assume only that the players do not suffer from strategic uncertainty. Player $i$ considers $\big(s_i,\Psi_i(s_i)\big)$ a Savage act. The set of all available Savage acts, i.e.,
\[
\mathcal{S}_i := \Big\{\big(s_i,\Psi_i(s_i)\big)\!:s_i\in A^\Omega_i~\text{is}~\sigma(\mathcal{I}_i)\text{-measurable}\Big\},
\]
is called the \textit{graph} of $\Psi_i$. Since the priors are equivalent, the players share the same $L^0$ space, i.e., a player considers two strategies identical if and only if each other player considers the same strategies identical.

Let $\Psi_{ij}(s_i)$ be the conjecture of Player $i$ about (the strategy of) Player $j$---given that Player $i$ applies the strategy $s_i$. The graph of $\Psi_{ij}$, i.e., $\mathcal{S}_{ij}$, is the set of all 2-tuples $\big(s_i,\Psi_{ij}(s_i)\big)$.

\begin{prop}\label{Pr.: Congruence}
If Assumption\/ \textup{\textbf{SAV}} is satisfied, $\mathcal{S}_{ij}$ and $\mathcal{S}_{ji}$ are congruent for $i,j=1,2,\ldots,n$.
\end{prop}

\noindent Proof: Suppose that $\mathcal{S}_{ij}$ and $\mathcal{S}_{ji}$ are incongruent for some $i\neq j$. Then there exists either a strategy $s_i$ such that $\Psi_{ji}\big(\Psi_{ij}(s_i)\big)\neq s_i$ or a strategy $s_j$ such that $\Psi_{ij}\big(\Psi_{ji}(s_j)\big)\neq s_j$. This means either $s_i$ would be unavailable to Player $i$ or $s_j$ would be unavailable to Player $j$.\QED

We conclude that $\mathcal{S}:=\mathcal{S}_1=\mathcal{S}_2=\ldots=\mathcal{S}_n$. More precisely, $\mathcal{S}$ is the set of Savage acts that are available to all players. The element of $\mathcal{S}$ that is manifested---\emph{ex post}---is said to be the solution of the game. All strategy tuples outside $\mathcal{S}$ cannot manifest. However, the solution of a game cannot be determined---\emph{ex ante}---without making any behavioral assumption.

\begin{theo}\label{Th.: Coherence principle}
Let Assumption\/ \textup{\textbf{SAV}} be satisfied. For all $s,t\in\mathcal{S}$ with $s\neq t$ we have that $s_i\neq t_i$ for $i=1,2,\ldots,n$.
\end{theo}

\noindent Proof: Proposition \ref{Pr.: Congruence} implies that $\Psi_{ij}$ is bijective for $i,j=1,2,\ldots,n$. This leads immediately to the statement of the theorem.\QED

Theorem \ref{Th.: Coherence principle} says that if a player changes his strategy, each other player must change his strategy, too. Hence, the players choose their strategies in a \emph{coherent} way. This ``coherence principle'' not only states that the strategic-independence assumption is violated in individual cases; it is \emph{always} violated under Assumption \textbf{SAV}. For this reason, we shall develop an appropriate notion of rationality in games without strategic uncertainty.

\begin{defn}[Rational solution]\label{Def.: Rational solution}
A solution $s\in\mathcal{S}$ is said to be\/ \emph{rational} if and only if\/ $\E_i\big(u_i(s)\big)\geq\E_i\big(u_i(t)\big)$ for all $t\in\mathcal{S}$ and $i=1,2,\ldots,n$.
\end{defn}

Hence, a rational solution is a Savage act that is optimal for all players. The set of rational solutions shall be denoted by $\mathcal{S}^*$. In general, a rational solution is \emph{not} a correlated equilibrium.

\begin{defn}[Pareto efficiency]
A solution $s\in\mathcal{S}$ is said to be\/ \emph{Pareto efficient} if and only if there is no $t\in\mathcal{S}$ such that $\E_i\big(u_i(t)\big)>\E_i\big(u_i(s)\big)$ and $\E_j\big(u_j(t)\big)\geq\E_j\big(u_j(s)\big)$ for each $j\neq i$.
\end{defn}

\begin{defn}[Uniqueness]
The elements of a set $\mathcal{R}\subseteq\mathcal{S}$ are said to be\/ \emph{essentially unique} if and only if\/ $\E_i\big(u_i(s)\big)=\E_i\big(u_i(t)\big)$ for all $s,t\in\mathcal{R}$ and $i=1,2,\ldots,n$.
\end{defn}

\begin{theo}\label{Th.: Solution}
Let Assumption\/ \textup{\textbf{SAV}} be satisfied. The rational solutions of a game are Pareto efficient and essentially unique.
\end{theo}

\noindent Proof: If $s\in\mathcal{S}^*$ is not Pareto efficient, there exists some $t\in\mathcal{S}$ such that $\E_i\big(u_i(t)\big)>\E_i\big(u_i(s)\big)$ and $\E_j\big(u_j(t)\big)\geq\E_j\big(u_j(s)\big)$ for each player $j\neq i$. The first inequality implies that $s$ cannot be a rational solution. Now, suppose that $s,t\in\mathcal{S}^*$ with $\E_i\big(u_i(s)\big)\neq\E_i\big(u_i(t)\big)$. Then either $\E_i\big(u_i(s)\big)<\E_i\big(u_i(t)\big)$ or $\E_i\big(u_i(s)\big)>\E_i\big(u_i(t)\big)$. Thus, Player $i$ can improve his expected utility by moving from $s$ to $t$ or from $t$ to $s$. This means either $s$ or $t$ cannot be a rational solution.\QED

In general we are not able to explain the way the solution of a strategic game takes place. Nonetheless, if we assume that the players are rational, we can find the possible sets of available Savage acts. The core idea can be illustrated this way: Suppose that we want to maximize some arbitrary function $\varphi\!:\R^2\rightarrow\R$ over its first argument. We could search for some point $x\in\R$ such that $\varphi(x,y)$ is maximal given that $y\in\R$ is fixed, but this approach fails if $y$ is an implicit function of $x$, i.e., $y=f(x)$. In this case, maximizing $\varphi$ requires us to maximize $\varphi\big(x,f(x)\big)$ over $x$. The two arguments $x$ and $y$ may be interpreted as strategies in a 2-person normal-form game. Further, the function $\varphi$ represents the expected utility of Player 1 and $f$ gives us the response of Player 2 to the strategy of Player 1, i.e., $f$ is the response function of Player 1. Let $\phi$ be the expected utility of Player 2 and $g$ his response function. Due to Proposition \ref{Pr.: Congruence} we have that $g=f^{-1}$. This means Player 2 has to maximize $\phi\big(f^{-1}(y),y\big)$ over $y$, whereas Player 1 maximizes $\varphi\big(x,f(x)\big)$ over $x$. A rational solution is a point $(x^*,y^*)\in\R^2$ that maximizes the expected utilities of both players, i.e.,
\[
\varphi\big(x^*,f(x^*)\big) \geq \varphi\big(x,f(x)\big)\qquad\text{and}\qquad
\phi\big(f^{-1}(y^*),y^*\big) \geq \phi\big(f^{-1}(y),y\big)
\]
for all $x,y\in\R$. Hence, if the players are rational, only those response functions that lead to a \emph{nonempty} set of rational solutions can be possible. This might substantially restrict the set of candidates for the response function $f$ and so we can deduce a meaningful set of Savage acts.

\begin{table}
  \centering
  \caption{Prisoner's Dilemma}\label{Tab.: Prisoner's Dilemma}
  \begin{tabular}{l}
  \toprule
  \begin{minipage}{12cm}
  \footnotesize{Two gangsters have committed a crime and have been arrested. Apart from illegal possession of arms, there is no evidence against them. Now, each prisoner can either deny or confess. If both prisoners deny, they serve only one year in prison. If one player denies and the other confesses, the latter is set free as a principal witness, whereas the former is sentenced to five years. Moreover, if both players confess, each of them serves four years in prison.}
  \end{minipage}\\
  \bottomrule
  \end{tabular}
\end{table}

This can easily be demonstrated by the prisoner's dilemma (see Table \ref{Tab.: Prisoner's Dilemma}). To make things as easy as possible, we may assume that $\Omega=\big\{0\big\}$, $\mathcal{F}=\big\{\emptyset,\Omega\big\}$, and $\mathcal{I}_1=\mathcal{I}_2=\big\{\Omega\big\}$. Further, the players possess the common action set $A=\big\{0,1\big\}$, where 0 means ``deny'' and 1 stands for ``confess.'' This structure is very simple and reflects perfect information. Indeed, we could also have chosen a more complicated model that allows for imperfect information, but this would not have any substantial impact on our conclusions.

Suppose that the prisoners are sitting together with their lawyer, who is instructed to announce their testimonies in court. He asks the prisoners to make their choices. In this case, each player is able to verify the strategy of the other. This means the players \emph{know} the other's response and so they choose their strategies coherently. This can be explained in a more formal way as follows: The players know everything that is determined by $\Omega$ and so they know the action of the other. Hence, the action of Player 2 is uniquely determined by the action of Player 1 and vice versa. For example, we could assume that Player 2 chooses Action 1 (``confess'') if Player 1 chooses Action 0 (``deny''). According to Theorem \ref{Th.: Coherence principle}, there must be a one-to-one correspondence between their actions. This means Player 1 must choose Action 0 if Player 2 chooses Action 1. Put another way, we have that $a_1=0\Leftrightarrow a_2=1$ and, equivalently, $a_1=1\Leftrightarrow a_2=0$. If Player 1 is rational he will choose $a_1=1\Rightarrow a_2=0$, but then Player 2 cannot be rational, since otherwise he would choose $a_2=1\Rightarrow a_1=0$, etc. Hence, the given response functions do not lead to a rational solution. This means we have that $\mathcal{S}^*=\emptyset$. By contrast, we could assume that $a_1=0\Leftrightarrow a_2=0$ and thus $a_1=1\Leftrightarrow a_2=1$.\footnote{This behavior is called ``tit for tat'' \citep[see, e.g.,][]{AH1981}.} Now, there exists a rational solution, i.e., both players deny. In fact, no player has an incentive to defect, since everybody \emph{knows} that the other player will then defect, too.\footnote{For this reason, ``confess'' does not dominate ``deny.''} Thus, we obtain a cooperative solution and not a correlated equilibrium.

Pareto efficiency and uniqueness are typical requirements in game theory \citep{Colman2004}. Theorem \ref{Th.: Solution} guarantees that these requirements are always satisfied. This is hardly surprising if the players can be \emph{certain} about each other's response. Indeed, such a situation indicates cooperative behavior. By contrast, noncooperative behavior requires strategic uncertainty.

\section{Strategic Uncertainty}

In most practical situations, players suffer from strategic uncertainty. This means they might not know each other's response. In the remainder of this note, I show that Aumann's formal approach can be applied to games with strategic uncertainty---after a slight but substantial modification: I drop Assumption \textbf{SAV}, i.e., I do no longer assume that the conjectures of the players are always correct. Hence, it might happen that $\Psi_i(s_i)\neq s^i$ for some strategy $s_i$.

Savage's theory of rational choice in games against nature can be maintained if we allow the decision maker to be wrong. This means he might not \emph{know} the outcome of his decision if some state of nature $\omega\in\Omega$ obtains, he must only \emph{believe} that $\omega$ leads to a specific consequence $s(\omega)$. We still assume that the state space, $\Omega$, is properly specified. More precisely, the decision maker has a \emph{unique} conjecture about nature's ``response'' to his strategy. Thus, he must be \emph{convinced} that the consequence $s(\omega)$ takes place if $\omega\in\Omega$ obtains, but his conjecture may be incorrect.

The same arguments can be applied to strategic games. We allow for strategic uncertainty and so the coherence principle need no longer hold true. Hence, players might give arbitrary responses to each other---according to their private information partitions and action sets. If a player has perfect information, he need not know the action of another player. ``Perfect'' just means that he is convinced about the actions of the other players. The point is that the player himself cannot distinguish between ``knowledge'' and ``conviction.'' In general, this is possible only for an outside observer. For this reason, our subjectivistic approach to rationality does not require knowledge. This allows us to explain strategic behavior that can be observed in everyday life in a very simple way.

Now, Player $i$ is considered rational if and only if he applies a strategy $s_i$ that leads to a Savage act $s=\big(s_i,\Psi_i(s_i)\big)\in\mathcal{S}_i$ such that $\E_i\big(u_i(s)\big)\geq\E_i\big(u_i(t)\big)$ for all $t\in\mathcal{S}_i$. If Assumption \textbf{FIX} is satisfied, we can substitute $\Psi_i(s_i)$ by $s^i_i$ for notational convenience. In this case, the solution of the game, $s^*=(s^*_1,s^*_2,\ldots,s^*_n)$, can be considered rational if and only if
\[
\E_i\Big(u_i\big(s^*_i,s^i_i\big)\Big) \geq \E_i\Big(u_i\big(s_i,s^i_i\big)\Big),\qquad\forall s_i,~i=1,2,\ldots,n\,.
\]
Moreover, if the players have a common prior, we can drop the index ``$i$'' from ``$\E_i$.'' Nonetheless, the given solution is not a subjective correlated equilibrium---unless we have that $s^i_i=s^{*i}$. This means if each player is rational \emph{and} the conjectures of all players are correct (for the given solution $s^*$) we obtain a subjective correlated equilibrium. However, since the players may suffer from strategic uncertainty, we refrain from making the additional assumption $\Psi_i(s^*_i)=s^{*i}$. This means the players can be rational although their conjectures about each other are incorrect.

Assumption \textbf{FIX} might be satisfied in most practical situations. Nonetheless, Rendezvous (see Table \ref{Tab.: Rendezvous}) shows that there exist strategic games in which it is more natural for a player to believe that the strategy of another player depends on his own strategy. Thus, it all depends on the epistemic conditions we are trying to reproduce by our model of strategic conflict. However, we do not have to explain \emph{how} a rational player comes to his conclusions. We can simply take his conjectures for granted---irrespective of their particular reasons.

For example, consider once again the prisoner's dilemma, but now suppose that there is no lawyer who represents the legal interests of the prisoners. We can still assume, for the sake of simplicity, that the players act on the basis of $\Omega$. This means they have perfect information. Nonetheless, the players suffer from strategic uncertainty and so neither \emph{knows} the action of the other. This might appear counterintuitive to the reader. As already mentioned, ``perfect'' means only that each player is convinced about the action of the other. However, their conjectures may be incorrect. Communication would not solve the basic problem, namely that the players are not able to verify the true intentions of the other---although they have perfect information. Irrespective of whatever one makes the other believe, each player remains ignorant of what the other is \emph{actually} going to do. Hence, everybody can defect without consequences. This implies that the players suffer from strategic uncertainty and we can readily justify Assumption \textbf{FIX}.

Now, it is always best for the players to confess. This leads to the well-known noncooperative solution of the prisoner's dilemma. Only if both players \emph{foresee} that the other is going to confess, we obtain a correlated equilibrium. For example, we could explain such a solution by assuming complete information and mutual knowledge of rationality. However, this goes beyond our subjectivistic approach to rational choice. We do not require the conjectures of the players to be correct---unless we want to drop the assumption of strategic uncertainty.

\section{Conclusion}

Aumann's model is based on the assumption that strategic uncertainty can always be translated into imperfect information and thus treated like uncertainty in a game against nature. Correlated equilibrium is guided by the idea that players are Bayes rational, have a common prior, and choose their strategies independently. We have seen that the Bayes condition is violated in every game with imperfect information. Further, if the players have perfect information, the action of a player always has an impact on the actions of the other players. More generally, if the players do not suffer from strategic uncertainty, they choose their strategies in a coherent way. This holds irrespective of whether the players are rational or not, but if they \emph{are} rational, the solution of the game is always Pareto efficient and essentially unique. Normally, since the players cannot choose their strategies independently, it is not a correlated equilibrium.

Hence, strategic independence requires strategic uncertainty. In most practical situations, players suffer from strategic uncertainty. Aumann's formal approach can be applied to games with strategic uncertainty if we allow the conjectures of the players to be incorrect. Each rational player chooses a strategy that maximizes his subjective expected utility given his conjectures about the other players. Nonetheless, even if the players are rational, have a common prior, and choose their strategies independently, the given solution need not be a correlated equilibrium---unless the conjectures of all players are correct in the particular case. Our subjectivistic approach to rational choice under strategic uncertainty does not require such an additional assumption and so the overall concept of equilibrium fades into the background.

\bibliographystyle{C:/Cloud/Local/LaTeX/BibTeX/mybib}
\bibliography{C:/Cloud/Local/LaTeX/BibTeX/mybib}

\end{document}